\RequirePackage{fix-cm}
\documentclass[twocolumn,epjc3]{svjour3_modified}  
\smartqed  % flush right qed marks, e.g. at end of proof
\RequirePackage{graphicx}
\RequirePackage[colorlinks,citecolor=blue,urlcolor=blue,linkcolor=blue]{hyperref}

\usepackage[acronym]{glossaries}
\usepackage{amsmath,amssymb,amsfonts}%
\usepackage{mathrsfs}%
\usepackage[title]{appendix}%
\usepackage{xcolor}%
\usepackage{textcomp}%
\usepackage{manyfoot}%
\usepackage{booktabs}%
\usepackage{algorithm}%
\usepackage{algorithmicx}%
\usepackage{algpseudocode}%
\usepackage{listings}%
\usepackage{subcaption}
\usepackage{lmodern}
\usepackage{lineno}
\usepackage{mathtools}
\usepackage{comment}
\journalname{Eur. Phys. J. C}
\usepackage{etoolbox}

\makeatletter
\patchcmd{\@maketitle}
  {\noindent{\small\@date\if@twocolumn\vskip 7.2mm\else\vskip 5.2mm\fi}}
  {\vskip 4mm
   \noindent{\small\@date}\par
   \vskip 5mm}
  {\typeout{EPJC date block adjusted successfully}}
  {\PackageWarning{EPJC}{Failed to adjust the date block}}
\makeatother

\begin{document}

\title{Dark Charge Conjugation and Complementary Constraints from
$D^0$ Mixing and Rare Decays with Massless Dark Photons%\thanksref{t1}
}
%\subtitle Do you have a subtitle?\\ If so, write it here}

\author{
X.~Zhong\textsuperscript{1,*} \and
X.~Q.~Shen\textsuperscript{1} 
}
\thankstext{e1}{e-mail: 15521018891@163.com}

\institute{{School of Physics and Mechanical and Electronical Engineering, Longyan University, Longyan 364012, China} }

\date{\today}

\maketitle

%\linenumbers

%abstract
\begin{abstract}

We investigate how dark charge conjugation organizes rare $D^0$ decays
involving a massless dark photon and how these decays complement neutral
charm mixing in constraining a common flavor changing interaction.
Within a simplified framework with a real pseudoscalar mediator,
$D^0\to\gamma_{\mathrm D}\gamma_{\mathrm D}$ probes the portal that
preserves $\mathcal{C}_{\mathrm D}$, whereas $D^0\to\gamma\gamma_{\mathrm D}$ probes the
portal associated with $\mathcal{C}_{\mathrm D}$ breaking.
Because the two decay channels and $D^0$ mixing share the same flavor
normalization and mediator propagator, the flavor dependence can be
eliminated analytically, converting the mixing constraint into a direct
upper bound on the invisible decay rate.
Combining current constraints from mixing, invisible decay, and the
single dark photon channel, we identify a transition between regions
dominated by mixing and regions dominated by invisible decay in mediator
parameter space.
The single dark photon search then yields conditional upper limits on
the portal associated with $\mathcal{C}_{\mathrm D}$ breaking once the flavor normalization
is fixed at its maximal value allowed by mixing and invisible decay.
These results demonstrate a symmetry resolved complementarity between
neutral meson mixing and rare charm decays in probing flavor changing
interactions with a massless dark sector.

\end{abstract}

\keywords{Dark photon, rare charm decays, neutral meson mixing, dark charge conjugation, flavor physics}

\section{Introduction} \label{intro}

An additional Abelian gauge symmetry provides a simple extension of the Standard Model (SM) and has motivated extensive studies of dark photon phenomenology \cite{holdom,pospelov,batell,bjorken}. A widely studied realization is a massive dark photon whose interactions with electrically charged matter arise through kinetic mixing \cite{batell,bjorken}. The massless case has a different theoretical structure. When the additional Abelian symmetry remains unbroken and
SM fields carry no dark charge, the representation of kinetic mixing
depends on the choice of gauge field basis, while physical observables
remain basis independent. Processes involving only SM external states
can then be insensitive to the mixing itself \cite{hoffmann,dobrescu,pan}.
Interactions of a massless dark photon with the visible sector can
instead arise through higher dimensional operators
~\cite{dobrescu,pan}.
 Such interactions have been studied in Higgs processes~\cite{gabrielli2014,biswas2016}, electroweak boson decays~\cite{fabbrichesi2018,cobal2020}, flavor changing fermion transitions~\cite{gabrielli2016}, and rare hadron decays~\cite{fabbrichesi2017,su2020kaon,su2020charm}. Flavor observables are of particular interest because they probe such interactions through transitions between SM quark flavors and through neutral meson mixing. This motivates the study of flavor systems in which decay and mixing observables can be examined within the same underlying flavor structure.

The neutral charm system provides a concrete realization of this possibility. A nonzero mass difference between the neutral charm mass eigenstates has been observed experimentally \cite{lhcb2021}. The SM interpretation of $D^0$ mixing remains affected by strong GIM suppression and sizable long distance hadronic contributions, which complicate a precise separation of possible new contributions from the SM amplitude \cite{lenz2021,falk2002,bobrowski2010,golowich2007}. This theoretical limitation motivates a phenomenological benchmark for the allowed size of an additional contribution to the mixing amplitude.
Rare charm decays provide an independent class of observables that probe the same underlying flavor transition. Belle has constrained invisible $D^0$ decays \cite{belle2017}, while BESIII has searched directly for $D^0 \to \gamma\gamma_{\mathrm D}$ and $D^0 \to \omega\gamma_{\mathrm D}$ \cite{besiii2025}. BESIII has also performed a related search in $\Lambda_c^+ \to p\gamma_{\mathrm D}$ \cite{besiii2022}. These experimental developments motivate a theoretical description in which rare charm decays and neutral charm mixing can be related to the same flavor changing interaction. 

The pseudoscalar mediator $a$ provides a simple low energy realization by linking the $c \to u$ transition to the electromagnetic and dark gauge sectors. Related flavor changing structures are widely studied in effective descriptions of light pseudoscalar states, where rare meson observables provide sensitive probes of the corresponding interactions \cite{izaguirre2017,bauer2017,bauer2021,bauer2022}. The relevant portal operators can be classified according to dark charge conjugation $\mathcal{C}_{\mathrm D}$. Because their transformation under $\mathcal{C}_{\mathrm D}$ depends on the number of dark photon fields, this symmetry provides a natural organization of the operators entering the decay amplitudes. The same $c \to u$ interaction also contributes to neutral charm mixing, linking the symmetry classification of the decay channels to an independent flavor constraint.

In this work, we investigate $D^0$ mixing, $D^0 \to \gamma_{\mathrm D}\gamma_{\mathrm D}$, and $D^0 \to \gamma\gamma_{\mathrm D}$ within this low energy framework. A common set of theoretical assumptions is applied to all three observables, with the flavor transition and mediator dynamics treated consistently. The analysis combines neutral charm mixing with the two decay channels while accounting for the different dependence of each observable on the portal interactions. Existing charm constraints are then applied to the corresponding flavor and portal parameters of the model.

\section{Simplified model and dark charge conjugation}
\label{sec:model}

We describe the process at the charm scale using a low energy
simplified theory containing a real pseudoscalar mediator \(a\) and a
massless dark photon gauge field \(A_{\mathrm D}^{\mu}\). The mediator
connects the flavor changing transition \(c\to u\) to either two dark
photons or one ordinary photon and one dark photon. Since this
connection is introduced through flavor changing effective operators
rather than ordinary kinetic mixing, the framework differs from the
minimal massive dark photon model. The relevant interactions are
\begin{equation}
\begin{aligned}
\mathcal{L}_{\mathrm{int}}={}&-a\!\left[y_{uc}(\mu)\,\bar{u}\,i\gamma_5 c
+y_{uc}^{*}(\mu)\,\bar{c}\,i\gamma_5 u\right]\\
&-\frac{g_{\gamma_{\mathrm D}\gamma_{\mathrm D}}}{4}\,aF_{\mathrm D}^{\mu\nu}\widetilde F_{{\mathrm D}\mu\nu}
-\frac{\kappa_{\gamma\gamma_{\mathrm D}}}{2}\,aF^{\mu\nu}\widetilde F_{{\mathrm D}\mu\nu},
\end{aligned}
\label{eq:interaction_lagrangian}
\end{equation}
where $F_{\mathrm D}^{\mu\nu}$ and $F^{\mu\nu}$ denote the dark
and electromagnetic field strength tensors, respectively. The dual tensor is defined by
$\widetilde X^{\mu\nu}\equiv
\epsilon^{\mu\nu\rho\sigma}X_{\rho\sigma}/2$,
with $\epsilon^{0123}=+1$.
The two gauge sector portal structures in Eq.~\eqref{eq:interaction_lagrangian} are related to those discussed in the dark axion portal framework~\cite{Kaneta:2016wvf}.
Here $\mu$ denotes the renormalization scale. The flavor coupling
$y_{uc}(\mu)$ is dimensionless and may be complex, whereas
$g_{\gamma_{\mathrm D}\gamma_{\mathrm D}}$ and
$\kappa_{\gamma\gamma_{\mathrm D}}$ have mass dimension $-1$.
The coupling $\kappa_{\gamma\gamma_{\mathrm D}}$ directly parametrizes
the dark charge conjugation breaking portal. The Lagrangian is therefore understood as a charm scale effective theory rather than a complete electroweak or ultraviolet construction.

The action of $\mathcal{C}_{\mathrm D}$ on the fields is specified by
\begin{equation}
\begin{aligned}
\mathcal{C}_{\mathrm D}:\qquad
&A_{\mathrm D}^\mu \to -A_{\mathrm D}^\mu,
\qquad
F_{\mathrm D}^{\mu\nu} \to -F_{\mathrm D}^{\mu\nu},\\
&a \to a,
\qquad
\text{SM fields} \to \text{SM fields}.
\end{aligned}
\label{eq:CD-transformation}
\end{equation}
The pseudoscalar mediator $a$ is taken to be even under this transformation, while all SM fields are invariant. It follows that $aF_{\mathrm D}\widetilde F_{\mathrm D}$ is even, whereas $aF\widetilde F_{\mathrm D}$ is odd. In the exact symmetry limit, $\kappa_{\gamma\gamma_{\mathrm D}}=0$, so amplitudes containing an odd number of dark photons vanish, while the two dark photon channel remains allowed. The corresponding amplitudes scale as
\begin{equation}
\begin{aligned}
\mathcal{M}(D^0\to\gamma_{\mathrm D}\gamma_{\mathrm D})
&\propto y_{uc}(\mu)g_{\gamma_{\mathrm D}\gamma_{\mathrm D}},\\
\mathcal{M}(D^0\to\gamma\gamma_{\mathrm D})
&\propto y_{uc}(\mu)\kappa_{\gamma\gamma_{\mathrm D}}.
\end{aligned}
\label{eq:amplitude_scaling}
\end{equation}
This selection rule connects the symmetry structure of the model to the relevant decay channels. Invisible decays probe the symmetry preserving portal, while final states with one dark photon probe the symmetry breaking portal. No interactions beyond those in Eq.~\eqref{eq:interaction_lagrangian} are included in the minimal benchmark.

To connect the quark level interaction to the physical meson, we introduce one common hadronic matrix element
\begin{equation}
H_{D^0}(\mu)
\equiv
\left|
\langle 0|\bar u\,i\gamma_5c|D^0\rangle
\right|
=
\frac{f_{D^0}m_{D^0}^2}
{m_c(\mu)+m_u(\mu)},
\label{eq:HD0-definition}
\end{equation}
where $m_{D^0}$ and $f_{D^0}$ denote the $D^0$ mass and decay
constant, respectively, while $m_c(\mu)$ and $m_u(\mu)$ are the running
quark masses. The quark masses, $y_{uc}(\mu)$, and the pseudoscalar density must be evaluated in the same renormalization scheme and at the same scale. The product $y_{uc}(\mu)H_{D^0}(\mu)$ then
supplies a consistent normalization for both rare decays
and $D^0$--$\bar D^0$ mixing. For notational simplicity, the common scale argument $\mu$ in
$y_{uc}(\mu)$ and $H_{D^0}(\mu)$ is suppressed below when no ambiguity arises. We retain
the phase through $y_{uc}=|y_{uc}|e^{i\phi_{uc}}$. The rare decay rates depend on $|y_{uc}|^2$, whereas the mixing
amplitude retains the phase through $y_{uc}^2$.

In the rare decays, the mediator carries the full meson momentum,
\(q^2=m_{D^0}^2\), so the relevant processes proceed through the virtual
transition
\(D^0\to a^*\to
\gamma_{\mathrm D}\gamma_{\mathrm D}\)
or
\(D^0\to a^*\to\gamma\gamma_{\mathrm D}\). Factoring out the conventional overall $i$, we define the reduced
mediator propagator in the constant width Breit--Wigner form,
with $s=q^2$ denoting the mediator virtuality,
\begin{equation}
\begin{aligned}
P_a(s)
&\equiv \frac{1}{s-m_a^2+i\Sigma_a},
\\
\Sigma_a
&=m_a\Gamma_a,
\\
\Gamma_a
&=\Gamma(a\to\gamma_{\mathrm D}\gamma_{\mathrm D})
 +\Gamma(a\to\gamma\gamma_{\mathrm D}),
\\
\Gamma(a\to\gamma_{\mathrm D}\gamma_{\mathrm D})
&=\frac{|g_{\gamma_{\mathrm D}\gamma_{\mathrm D}}|^2m_a^3}{64\pi},
\\
\Gamma(a\to\gamma\gamma_{\mathrm D})
&=\frac{|\kappa_{\gamma\gamma_{\mathrm D}}|^2m_a^3}{32\pi}.
\end{aligned}
\label{eq:propagator-and-width}
\end{equation}
Here $m_a$ and $\Gamma_a$ denote the mediator mass and total width,
respectively.
The width is evaluated point by point from
Eq.~\eqref{eq:propagator-and-width}, while it is treated as independent of $s$ in the
propagator. Flavor induced hadronic contributions associated with $y_{uc}$
are not included in this prescription, since a dedicated treatment
at the charm scale would require additional hadronic input.
No additional independent mediator decay channels are included
in the baseline prescription.

Within this framework, neutral meson mixing constrains \(y_{uc}\), the
invisible decay bound constrains
\(y_{uc}g_{\gamma_{\mathrm D}\gamma_{\mathrm D}}\), and the
single dark photon search constrains
\(y_{uc}\kappa_{\gamma\gamma_{\mathrm D}}\). The corresponding observables are formulated using the
common hadronic matrix element, phase convention, and mediator
propagator introduced above.

\section{Observable quantities}
\label{observable_quantities}

Building on the interaction Lagrangian, hadronic matrix element, and
propagator defined in Sect.~\ref{sec:model}, we now derive the decay
and mixing observables relevant to the experimental constraints. For
later convenience, we introduce the squared inverse propagator
\begin{equation}
\mathcal R_a(s)
\equiv
\left|\mathcal P_a(s)\right|^{-2}
=
\left(s-m_a^2\right)^2+\Sigma_a^2.
\label{eq:propagator-denominator}
\end{equation}
The hadronic matrix element \(H_{D^0}\) is defined in
Eq.~\eqref{eq:HD0-definition}. Far from resonance,
\(\mathcal R_a(m_{D^0}^2)\) approaches
\((m_{D^0}^2-m_a^2)^2\), while near resonance the finite width
contribution prevents an artificial divergence and must be retained.

The invisible decay width and the corresponding branching fraction are
\begin{equation}
\begin{aligned}
\Gamma_{\gamma_{\mathrm D}\gamma_{\mathrm D}}
&\equiv \Gamma(D^0\to\gamma_{\mathrm D}\gamma_{\mathrm D}) \\
&=
\frac{
|y_{uc}|^2 H_{D^0}^2
|g_{\gamma_{\mathrm D}\gamma_{\mathrm D}}|^2
m_{D^0}^3
}{
64\pi\mathcal R_a(m_{D^0}^2)
},
\\
\mathcal B_{\gamma_{\mathrm D}\gamma_{\mathrm D}}
&=
\frac{
\Gamma_{\gamma_{\mathrm D}\gamma_{\mathrm D}}
}{
\Gamma_{D^0}^{\mathrm{tot}}
},
\end{aligned}
\label{eq:invisible-width}
\end{equation}
where $\Gamma_{D^0}^{\rm tot}$ denotes the total $D^0$ width and
$\tau_{D^0}$ its lifetime, with
$\Gamma_{D^0}^{\rm tot}=\hbar/\tau_{D^0}$.
The factor $1/2!$ associated with the two identical dark photons
is included in Eq.~\eqref{eq:invisible-width}. The same meson to mediator transition
also contributes to mixing between $D^0$ and its antiparticle.
In the single pole approximation, and using the standard
neutral $D$ meson mixing convention~\cite{lenz2021}, the complex
mediator contribution is
\begin{equation}
\begin{alignedat}{2}
\mathcal M_{12}^{(a)}
&\equiv
M_{12}^{(a)}
-\frac{i}{2}\Gamma_{12}^{(a)}
&&=
\frac{y_{uc}^{2}H_{D^0}^{2}}{2m_{D^0}}
\mathcal P_a(m_{D^0}^{2}),
\\[4pt]
&&
\mathllap{2\left|\mathcal M_{12}^{(a)}\right|}
&\leq
\Delta m_{D^0,\mathrm{NP}}^{\max}.
\end{alignedat}
\label{eq:mixing-constraint}
\end{equation}
Here $M_{12}^{(a)}$ and $\Gamma_{12}^{(a)}$ denote the
dispersive and absorptive parts of the mediator contribution
to $D^0$--$\bar D^0$ mixing, respectively. The subscript
$\mathrm{NP}$ identifies the mediator induced contribution beyond
the Standard Model, while $\Delta m_{D^0,\mathrm{NP}}^{\max}$
denotes the corresponding benchmark upper value.
Its numerical prescription is specified in Sect.~\ref{numerical_analysis}.

The second relation in Eq.~\eqref{eq:mixing-constraint} defines the phase independent
no cancellation benchmark adopted in this analysis. It is not
interpreted as an exact relation between the measured mass
splitting and the full complex mixing amplitude. The consistency region introduced below excludes the immediate
pole region, where absorptive effects can become important.
A treatment including possible cancellations and the complete
dependence on the mass and width differences and the mixing phase
is beyond the benchmark considered here.

When dark charge conjugation is broken, the mixed portal produces an
ordinary photon and a dark photon. Since the two final state particles
are distinguishable, no identical particle factor is required. Using
Eq.~\eqref{eq:invisible-width}, the corresponding decay width can be
written as
\begin{equation}
\begin{aligned}
\Gamma_{\gamma\gamma_{\mathrm D}}
&\equiv
\Gamma(D^0\to\gamma\gamma_{\mathrm D})
\\
&=
\frac{|y_{uc}|^2 H_{D^0}^2
|\kappa_{\gamma\gamma_{\mathrm D}}|^2
m_{D^0}^3}
{32\pi R_a(m_{D^0}^2)}
\\
&=
2\left|
\frac{\kappa_{\gamma\gamma_{\mathrm D}}}
{g_{\gamma_{\mathrm D}\gamma_{\mathrm D}}}
\right|^2
\Gamma_{\gamma_{\mathrm D}\gamma_{\mathrm D}},
\qquad
g_{\gamma_{\mathrm D}\gamma_{\mathrm D}}\neq0.
\end{aligned}
\label{eq:single-dark-photon-width}
\end{equation}
The two channels contain the same flavor changing coupling, hadronic
matrix element, and mediator propagator. Their relative rate therefore
depends only on the two portal couplings. Accordingly, the single dark photon width vanishes when
$\kappa_{\gamma\gamma_{\mathrm D}}=0$.

The common origin of mixing and decay allows the factor
\(|y_{uc}|^2H_{D^0}^2\) to be eliminated. Using the mixing constraint in
Eq.~\eqref{eq:mixing-constraint} gives
\begin{equation}
\begin{aligned}
\mathcal B_{\gamma_{\mathrm D}\gamma_{\mathrm D}}^{\max,\mathrm{mix}}
&=
\frac{
\Delta m_{D^0,\mathrm{NP}}^{\max}
}{
\Gamma_{D^0}^{\mathrm{tot}}
}
\frac{
|g_{\gamma_{\mathrm D}\gamma_{\mathrm D}}|^2m_{D^0}^4
}{
64\pi\sqrt{\mathcal R_a(m_{D^0}^2)}
},
\\
\left|
\frac{\kappa_{\gamma\gamma_{\mathrm D}}}{g_{\gamma_{\mathrm D}\gamma_{\mathrm D}}}
\right|
&\leq
\left[
\frac{
\mathcal B_{\gamma\gamma_{\mathrm D}}^{\mathrm{lim}}
}{
2\mathcal B_{\gamma_{\mathrm D}\gamma_{\mathrm D}}^{\mathrm{pred}}
}
\right]^{1/2}.
\end{aligned}
\label{eq:joint-constraints}
\end{equation}
The quantity $\mathcal B_{\gamma_{\mathrm D}\gamma_{\mathrm D}}^{\max,\mathrm{mix}}$
denotes the maximal double dark photon branching fraction allowed
by the mixing constraint.
The quantity $\mathcal B_{\gamma\gamma_{\mathrm D}}^{\lim}$ denotes the
experimental upper limit on the single dark photon branching
fraction, while
$\mathcal B_{\gamma_{\mathrm D}\gamma_{\mathrm D}}^{\mathrm{pred}}$ denotes the
predicted double dark photon branching fraction at the parameter
point under consideration. The first relation quantifies the
complementarity between mixing and invisible decay searches.
The second constrains the relative strength of the
symmetry breaking portal. When the mixed portal contributes to
$\Gamma_a$, the second relation is implicit because the predicted
branching fraction depends on the mediator width. It is therefore
evaluated self-consistently using Eq.~\eqref{eq:propagator-and-width}. At $g_{\gamma_{\mathrm D}\gamma_{\mathrm D}}=0$, the constraint is evaluated
directly from the first expression from Eq.~\eqref{eq:single-dark-photon-width}.

The expressions satisfy the expected dimensional and symmetry limits.
Both decay widths have mass dimension one and all branching fractions
are dimensionless. The single dark photon channel vanishes as
$\kappa_{\gamma\gamma_{\mathrm D}}\to0$, while the double dark photon
channel remains finite. Away from resonance, the zero width
approximation is recovered, whereas near resonance the same finite width
prescription must be used consistently in both decay and mixing.

\section{Numerical analysis and phenomenological constraints}
\label{numerical_analysis}

The numerical analysis combines the constraints from neutral meson mixing, invisible decay, and the single dark photon channel. The $D^0$ mass, lifetime, mass splitting, and the relevant branching fraction limits are taken from the 2026 Review of Particle Physics (PDG)~\cite{PDG2026}. The inputs used in the figures are $m_{D^0}=1.86484\,\mathrm{GeV}$, $\tau_{D^0}=4.103\times10^{-13}\,\mathrm{s}$, $\Delta m_{D^0}=(98\pm11)\times10^8\,\hbar\,\mathrm{s}^{-1}$, $\mathcal B(D^0\to\mathrm{invisible})<9.4\times10^{-5}$, and $\mathcal B(D^0\to\gamma\gamma_{\mathrm D})<2.0\times10^{-6}$. Both branching fraction limits are imposed at the 90\% confidence
level. For the no cancellation benchmark, we define the allowed mediator contribution using the PDG 2026 ``OUR AVERAGE'' value of $\Delta m_{D^0}$. Treating the quoted uncertainty
as Gaussian, we take the corresponding one sided 90\% upper value,
which yields
$\Delta m_{D^0,\mathrm{NP}}^{\max}
=7.38\times10^{-15}\,\mathrm{GeV}$.
The mediator width is evaluated according to Eq.~\eqref{eq:propagator-and-width}. For $\kappa_{\gamma\gamma_{\mathrm D}}=0$, only the
$a\to\gamma_{\mathrm D}\gamma_{\mathrm D}$ contribution remains in the baseline
radiative width. When the mixed portal is present, the $a\to\gamma\gamma_{\mathrm D}$ contribution is
included consistently in $\Gamma_a$.
A partonic estimate using the maximal flavor coupling allowed by mixing
shows that flavor-induced contributions have a negligible effect on the
mediator propagator over the benchmark region examined in~\ref{app:flavor_width}.

To retain a narrow mediator and avoid
results dominated by the pole region, we require
$\Gamma_a/m_a<0.1$ and $|m_a-m_{D^0}|>5\Gamma_a$.
These conditions define the consistency region used throughout the numerical analysis. In the numerical illustrations, we focus on the high mass side of the $D^0$ pole and consider
$1.90 \leq m_a \leq 3.20~\mathrm{GeV}$,
which covers the transition and closure of the shaded region within the adopted consistency domain.

\begin{figure}[t]
\centering\includegraphics[width=\columnwidth]{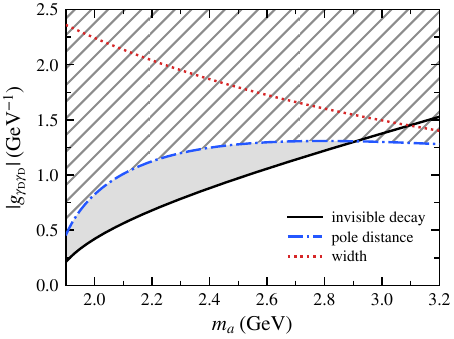}
  \caption{Combined constraints from neutral meson mixing and invisible
decay in the $(m_a,|g_{\gamma_{\mathrm D}\gamma_{\mathrm D}}|)$ plane for $\kappa_{\gamma\gamma_{\mathrm D}}=0$. The shaded region indicates where invisible decay gives the stronger bound on $|y_{uc}|$. The hatched region lies outside the adopted consistency domain. The solid, dash dotted, and dotted curves show the invisible decay, pole distance, and width boundaries.}
  \label{fig:mediator_plane}
\end{figure}

\begin{figure}[t]
  \centering
  \includegraphics[width=\columnwidth]{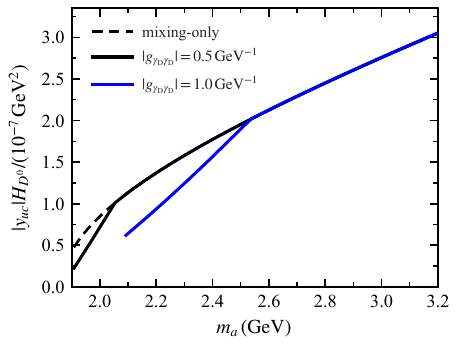}
  \caption{Upper bounds on $|y_{uc}|H_{D^0}$ for
$|g_{\gamma_{\mathrm D}\gamma_{\mathrm D}}|=0.5$ and
$1.0\,\mathrm{GeV}^{-1}$ with
$\kappa_{\gamma\gamma_{\mathrm D}}=0$.
The dashed curve shows the mixing-only limit, while the black (blue)
solid curve shows the combined mixing--invisible-decay limit for the
smaller (larger) portal coupling. Only points inside the
consistency domain are displayed.}
  \label{fig:flavor_normalization}
\end{figure}

Figure~\ref{fig:mediator_plane} shows how the mixing and invisible decay constraints combine in the $(m_a,|g_{\gamma_{\mathrm D}\gamma_{\mathrm D}}|)$ plane. At each point, the mixing bound fixes the largest allowed double dark photon branching fraction. Comparing this value with the invisible decay limit determines whether the latter further reduces the allowed range of $|y_{uc}|$. The shaded region marks the parameter domain in which this additional restriction occurs. It is not a direct exclusion of the mediator point because smaller values of $|y_{uc}|$ remain allowed. The hatched region fails at least one of the width or pole distance requirements. As $m_a$ increases, the invisible decay boundary moves toward larger $|g_{\gamma_{\mathrm D}\gamma_{\mathrm D}}|$ because a stronger portal coupling is needed to compensate for the increasing propagator suppression. The shaded region remains connected and closes only when it reaches the consistency boundary. The complementarity is therefore supported by a finite parameter domain rather than by an isolated enhancement near the pole.

Figure~\ref{fig:flavor_normalization}  expresses the same comparison in terms of the flavor
normalization $|y_{uc}|H_{D^0}$ for the representative benchmarks
$|g_{\gamma_{\mathrm D}\gamma_{\mathrm D}}|=0.5$ and
$1.0\,\mathrm{GeV}^{-1}$. As in Figure~\ref{fig:mediator_plane}, the exact
$\mathcal{C}_{\mathrm D}$-symmetric limit, $\kappa_{\gamma\gamma_{\mathrm D}}=0$, is
adopted. These benchmarks are chosen such that the mixing-invisible
crossovers occur at well separated mediator masses within the
adopted consistency domain. The mixing-only limits for the two benchmarks nearly overlap and appear as
a single dashed curve. For the smaller portal coupling, the invisible decay
constraint becomes relevant at lower mediator masses. Increasing
$|g_{\gamma_{\mathrm D}\gamma_{\mathrm D}}|$ shifts the affected interval
toward larger $m_a$, since the stronger portal coupling compensates for the
increasing propagator suppression. Each solid curve merges with the
mixing-only limit once mixing again provides the stronger constraint.

\begin{figure}[t]
  \centering
  \includegraphics[width=\columnwidth]{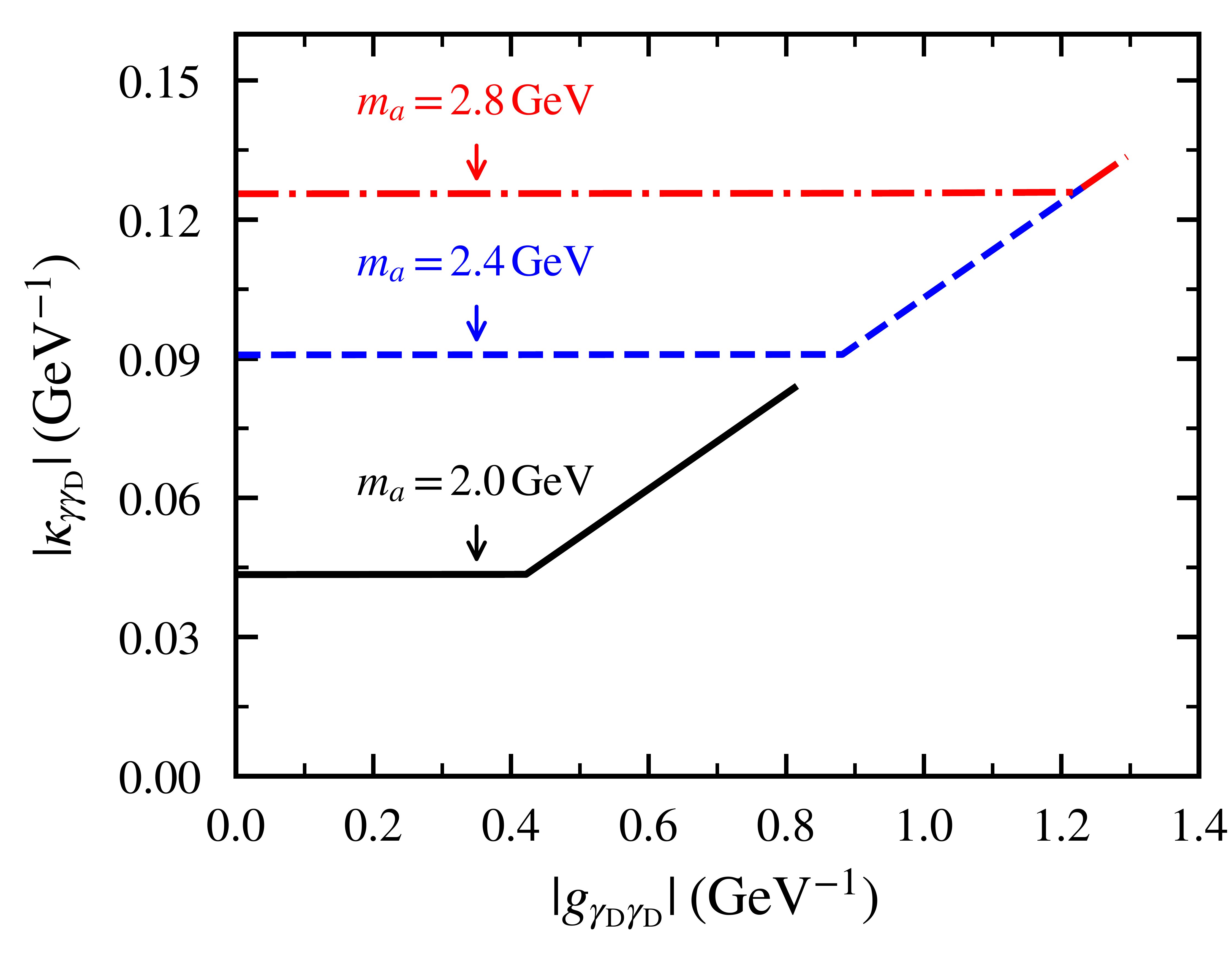}
  \caption{Conditional upper envelopes for
$|\kappa_{\gamma\gamma_{\mathrm D}}|$ at
$m_a=2.0,\ 2.4,\ \text{and}\ 2.8~\mathrm{GeV}$.
The flavor normalization saturates the combined mixing and invisible decay
bound, with the mediator width and propagator treated self-consistently.}
  \label{fig:portal_plane}
\end{figure}

The single dark photon constraint is then mapped onto the two portal couplings
in Figure~\ref{fig:portal_plane} for three representative mediator masses within the numerical scan. For each $m_a$ and
$|g_{\gamma_{\mathrm D}\gamma_{\mathrm D}}|$, the mediator width and propagator are
updated consistently as $|\kappa_{\gamma\gamma_{\mathrm D}}|$ is varied,
while the flavor normalization is set to its largest value
permitted by mixing and invisible decay. With the flavor normalization fixed at its maximal allowed value, each curve gives the conditional upper limit on the mixed portal coupling compatible with the single dark photon limit and the
consistency conditions. At small values
of the even portal coupling, mixing controls the maximal double dark photon
rate and the envelope is nearly horizontal. Once invisible decay becomes
the stronger restriction, the envelope rises with increasing even portal
strength. Each curve terminates at the consistency boundary. A smaller
flavor coupling would correspondingly weaken the constraint on the mixed
portal coupling.

Taken together, the three figures show how the experimental constraints
are mapped onto the mediator, flavor, and portal parameter spaces.
They illustrate the transition between mixing and invisible decay dominated
regimes and the resulting constraint on the symmetry breaking portal.
These results highlight the complementary roles of neutral meson mixing,
invisible decay, and the single dark photon search.

\section{Conclusion}

We have studied rare $D^0$ decays into massless dark photons
within a simplified framework containing a real pseudoscalar
mediator. Dark charge conjugation provides a direct connection
between the symmetry structure of the model and the relevant
experimental observables. The decay
$D^0\to\gamma_{\mathrm D}\gamma_{\mathrm D}$ probes the portal that remains allowed
in the exact $\mathcal{C}_{\mathrm D}$ limit, whereas
$D^0\to\gamma\gamma_{\mathrm D}$ probes the portal associated with
$\mathcal{C}_{\mathrm D}$ breaking. Neutral meson mixing constrains the common
flavor coupling that enters both decay channels. These observables
therefore probe different components of the same underlying
interaction structure.

Using current limits on invisible and single dark photon
$D^0$ decays together with the measured neutral meson mass
splitting, we have determined the resulting constraints on the
flavor and portal couplings. The analysis shows that mixing and
invisible decay constrain complementary regions of parameter
space. At smaller values of the even portal coupling, mixing
sets the dominant restriction on the allowed flavor normalization.
As the portal strength increases, the invisible decay constraint
can become more restrictive. The single dark photon search then
constrains the $\mathcal{C}_{\mathrm D}$ breaking portal and provides conditional
upper envelopes once the flavor normalization is fixed by mixing
and invisible decay. The mediator width and propagator are updated
consistently throughout this analysis when the mixed portal is
present.

The numerical results are obtained within a constant width
prescription and the phase independent no cancellation benchmark.
The immediate pole region is excluded by the adopted consistency
conditions, since a more complete treatment there would require
the full dependence on dispersive and absorptive mixing effects.
Possible cancellations in the mixing amplitude or additional
mediator decay channels could modify the numerical boundaries.
The flavor-induced contribution is numerically negligible in the benchmark
estimate. Within these assumptions,
the combined analysis demonstrates how neutral meson mixing,
invisible decay, and single dark photon searches can provide
complementary probes of flavor changing interactions with a
massless dark sector.

\begin{acknowledgements}
This study was supported by the Fujian Province Young and Middle-aged Scientists Fund (grants JAT241139), the Fujian Province Natural Science Fund General Project (grant 2025J011710), and the Longyan University Doctoral Research Startup Projects 2025 (grants LB2025003).
\end{acknowledgements}

\appendix
\section{Estimate of the flavor-induced mediator width}
\label{app:flavor_width}

We briefly check whether the omitted flavor induced mediator
width can affect the numerical analysis. From the mixing bound in
Eq.~\eqref{eq:mixing-constraint},
\begin{equation}
|y_{uc}|^2
\leq
\frac{\Delta m_{D^0,\mathrm{NP}}^{\max}m_{D^0}
\sqrt{\mathcal R_a(m_{D^0}^2)}}
{H_{D^0}^2}.
\label{eq:app_yuc_bound}
\end{equation}
At parton level, and neglecting $m_u$, we denote the combined width
for $a\to c\bar u$ and $a\to \bar c u$ by
$\Gamma_{uc}^{\mathrm{part}}$. For this estimate, the radiative width
defined in Eq.~\eqref{eq:propagator-and-width} is denoted by $\Gamma_a^{\mathrm{rad}}$. The two
quantities are
\begin{equation}
\begin{aligned}
\Gamma_{uc}^{\rm part}
&\simeq
\frac{3|y_{uc}|^2m_a}{4\pi}
\left(1-\frac{m_c^2}{m_a^2}\right)^2,
\\
\Gamma_a^{\rm rad}
&=
\frac{m_a^3}{64\pi}
\left(
|g_{\gamma_{\mathrm D}\gamma_{\mathrm D}}|^2
+2|\kappa_{\gamma\gamma_{\mathrm D}}|^2
\right).
\end{aligned}
\label{eq:app_widths}
\end{equation}

For the numerical estimate, we take
\linebreak
$m_c(m_c)=1.2729~\mathrm{GeV}$ and
$f_{D^0}=0.2120~\mathrm{GeV}$. Since $m_u\ll m_c$ at the scale considered, we neglect $m_u$
in this order of magnitude estimate.
Eq.~\eqref{eq:HD0-definition} then gives
$H_{D^0}(m_c)=0.579~\mathrm{GeV}^2$.
The running flavor quantities entering this estimate are evaluated
at the common scale $\mu=m_c$. We scan $1.90\leq m_a\leq3.20~\mathrm{GeV}$
with $|g_{\gamma_{\mathrm D}\gamma_{\mathrm D}}|\geq0.05~\mathrm{GeV}^{-1}$. For a conservative estimate, $|y_{uc}|$ is set to its maximal
value allowed by Eq.~(\ref{eq:app_yuc_bound}) and
$\kappa_{\gamma\gamma_{\mathrm D}}=0$. The largest ratio occurs at the
upper end of the scan, $m_a=3.20~\mathrm{GeV}$, where we find
$\Gamma_{uc}^{\rm part}/\Gamma_a^{\rm rad}
\lesssim4\times10^{-10}$.
Thus the flavor-induced contribution has a negligible effect on the
mediator propagator over the parameter range tested here. The partonic estimate is used
only as a robustness check and is not intended as a precision
description of charm scale hadronic decays.

%\bibliography{darkphoton_paper}   % name your BibTeX data base

\end{document}